\documentclass[letterpaper]{article} % DO NOT CHANGE THIS
\usepackage{aaai2026}
\usepackage{times}  % DO NOT CHANGE THIS
\usepackage{helvet}  % DO NOT CHANGE THIS
\usepackage{courier}  % DO NOT CHANGE THIS
\usepackage[hyphens]{url}  % DO NOT CHANGE THIS
\usepackage{graphicx} % DO NOT CHANGE THIS
\usepackage{natbib}  % DO NOT CHANGE THIS AND DO NOT ADD ANY OPTIONS TO IT
\usepackage{caption} % DO NOT CHANGE THIS AND DO NOT ADD ANY OPTIONS TO IT
\usepackage{algorithm}
\usepackage{algorithmic}

\usepackage{newfloat}
\usepackage{listings}
\DeclareCaptionStyle{ruled}{labelfont=normalfont,labelsep=colon,strut=off} % DO NOT CHANGE THIS
\floatstyle{ruled}
\newfloat{listing}{tb}{lst}{}
\floatname{listing}{Listing}
\usepackage{amsmath}
\usepackage{booktabs}
\usepackage{multirow}
\usepackage{subfigure}

\usepackage{color} % (if used in text)
\nocopyright
\usepackage{hyperref}

\title{PASE: Leveraging the Phonological Prior of WavLM \\for Low-Hallucination Generative Speech Enhancement}

\author{
    Xiaobin Rong\textsuperscript{\rm 1,2},
    Qinwen Hu\textsuperscript{\rm 1,2},
    Mansur Yesilbursa\textsuperscript{\rm 2},
    Kamil Wojcicki\textsuperscript{\rm 2},
    Jing Lu\textsuperscript{\rm 1}
}
\affiliations{
    \textsuperscript{\rm 1}Key Laboratory of Modern Acoustics, Nanjing University\\
    \textsuperscript{\rm 2}Collaboration AI, Cisco Systems, Inc.\\
    \href{mailto:xiaobin.rong@smail.nju.edu.cn}{xiaobin.rong@smail.nju.edu.cn}, \href{mailto:lujing@nju.edu.cn}{lujing@nju.edu.cn}
}

\usepackage{bibentry}
\begin{document}

\maketitle

\begin{abstract}
Generative models have shown remarkable performance in speech enhancement (SE), achieving superior perceptual quality over traditional discriminative approaches. However, existing generative SE approaches often overlook the risk of hallucination under severe noise, leading to incorrect spoken content or inconsistent speaker characteristics, which we term linguistic and acoustic hallucinations, respectively.
We argue that linguistic hallucination stems from models' failure to constrain valid phonological structures and it is a more fundamental challenge.
While language models (LMs) are well-suited for capturing the underlying speech structure through modeling the distribution of discrete tokens, existing approaches are limited in learning from noise-corrupted representations, which can lead to contaminated priors and hallucinations.
To overcome these limitations, we propose the Phonologically Anchored Speech Enhancer (PASE), a generative SE framework that leverages the robust phonological prior embedded in the pre-trained WavLM model to mitigate hallucinations. 
First, we adapt WavLM into a denoising expert via representation distillation to clean its final-layer features. Guided by the model's intrinsic phonological prior, this process enables robust denoising while minimizing linguistic hallucinations.
To further reduce acoustic hallucinations, we train the vocoder with a dual-stream representation: the high-level phonetic representation provides clean linguistic content, while a low-level acoustic representation retains speaker identity and prosody.
Experimental results demonstrate that PASE not only surpasses state-of-the-art discriminative models in perceptual quality, but also significantly outperforms prior generative models with substantially lower linguistic and acoustic hallucinations.
\end{abstract}

\begin{links}
    \link{Code}{https://github.com/cisco-open/pase}
    \link{Demo}{https://xiaobin-rong.github.io/pase_demo/}
\end{links}
% Uncomment the following to link to your code, datasets, an extended version or similar.
% You must keep this block between (not within) the abstract and the main body of the paper.
% \begin{links}
%     \link{Code}{https://aaai.org/example/code}
%     \link{Datasets}{https://aaai.org/example/datasets}
%     \link{Extended version}{https://aaai.org/example/extended-version}
% \end{links}

\section{Introduction}
Speech enhancement (SE) aims to recover clean speech from noisy mixtures, improving both quality and intelligibility. Deep learning-based SE methods can be broadly categorized into discriminative and generative approaches. 
While the former is effective at noise reduction, it often struggles to preserve speech naturalness under challenging conditions \cite{FlowSE}. To overcome this, generative models have recently emerged as a compelling alternative, demonstrating strong capabilities in synthesizing speech with superior perceptual quality \cite{CDiffSE}. These models learn the distribution of clean speech using techniques such as generative adversarial networks (GANs) \cite{Efficient_SSL_SE}, diffusion models \cite{CDiffSE, StoRM}, flow matching \cite{FlowSE}, and language models (LMs) \cite{SELM, Genhancer}. 

Despite their strengths, however, generative models are prone to hallucinations, which could result in inconsistencies in linguistic content or speaker characteristics between the enhanced and original noisy speech \cite{URGENT2}. This problem was initially overlooked as it cannot be effectively detected by commonly used non-intrusive quality metrics \cite{Eval_generative}, such as DNSMOS \cite{DNSMOS, DNSMOS-P835} and UTMOS \cite{UTMOS}. 

While recent studies \cite{FlowSE, Genhancer, LLaSE-G1} have started to incorporate hallucination-sensitive metrics such as word error rate (WER) and speaker similarity (SpkSim), these evaluations are mostly conducted under high signal-to-noise ratio (SNR) conditions, where models tend to perform reliably. However, hallucinations become much more severe in low-SNR conditions, potentially compromising the system's practicality. For instance, GenSE \cite{GenSE} achieves strong performance on high-SNR test sets; however, it still reports a WER as high as 28.4\% on a more challenging test set, indicating substantial room for improvement in mitigating hallucinations under adverse conditions.

In this paper, we aim to mitigate hallucinations in generative speech enhancement. We begin by categorizing hallucinations into two disentangled aspects: \textit{linguistic} and \textit{acoustic}, based on whether the distortion affects speech content or speaker characteristics. We argue that acoustic hallucination, which arises from the loss of fine-grained details, can be alleviated by supplying acoustic cues. In contrast, linguistic hallucination, stemming from models' failure to constrain valid phonological structures, is the more fundamental challenge. While existing LM-based approaches are well-suited to capturing speech structure through token-based language modeling, we contend that they are fundamentally limited in two ways: (1) they learn phonological priors from noise-corrupted input representations, which risks leading to unreliable or contaminated knowledge; and (2) their reliance on a limited number of discrete tokens inherently discards essential acoustic information such as pitch and timbre, making them ill-equipped to handle acoustic hallucinations.

To overcome these limitations, we propose the \textbf{P}honologically \textbf{A}nchored \textbf{S}peech \textbf{E}nhancer (PASE), a novel generative framework that circumvents the pitfalls of LM-based approaches. First, instead of learning a phonological prior from corrupted inputs, PASE directly leverages the robust, pre-existing prior embedded in the self-supervised model WavLM \cite{WavLM}. Second, PASE operates entirely in the continuous representation space, avoiding the information loss from discretization that contributes to acoustic hallucinations.
Specifically, we first adapt WavLM into a denoising expert via representation distillation. This process is guided by WavLM's intrinsic phonological prior, enabling effective denoising while %resisting linguistic hallucinations. 
being robust to linguistic hallucinations.
Subsequently, a vocoder is trained to reconstruct the waveform with a dual-stream representation. Following established layer-wise analyses of WavLM \cite{Comparative_layerwise_SSL_analysis}, we use phonetic representations from the final transformer layer to ensure linguistic integrity, while conditioning on acoustic representations from the first layer to preserve speaker characteristics.
Our contributions are three-fold:

\begin{itemize}
    \item \textbf{A New Conceptualization for Hallucination Analysis}: 
    We are, to the best of our knowledge, the first in speech enhancement to formally categorize hallucinations into linguistic and acoustic types and trace them to their distinct root causes, offering a new lens for analyzing the fundamental flaws of existing generative SE paradigms.
    \item \textbf{A Novel Prior-Leveraging Paradigm}: We propose PASE, a framework that marks a paradigm shift from learning potentially flawed knowledge to directly leveraging robust and pre-existing phonological prior, thereby effectively alleviating linguistic hallucinations, while a dual-stream, acoustic-conditioned design simultaneously mitigates acoustic hallucinations.
    \item \textbf{State-of-the-Art Performance}: PASE establishes new state-of-the-art (SOTA) results through extensive experiments, significantly outperforming existing generative and discriminative models in reducing both linguistic and acoustic hallucinations, while delivering superior perceptual quality with lower computational complexity.
\end{itemize}

\section{Related Works}
\subsection{Knowledge Priors in Self-Supervised Speech Models}
Self-supervised speech models (S3Ms) learn powerful representations from large-scale unlabeled data by employing self-supervised learning (SSL) objectives, such as masked prediction \cite{HuBERT, WavLM} and auto-regressive prediction \cite{SSL_AR}. These pre-training paradigms force the model to capture the underlying structure of speech, resulting in a hierarchy of representations where different layers encode distinct aspects of information \cite{Layerwise_SSL_analysis}. A consistent finding is an acoustic-to-linguistic progression: lower layers capture fine-grained acoustic details and speaker characteristics, while higher layers encode more abstract linguistic properties ostensibly related to phonetics, syntax, and even semantics \cite{Phonetic_analysis, Phonology_analysis, Syntax_analysis, Semantic_analysis}.

However, the true nature of this high-level ``linguistic" knowledge requires careful scrutiny. A key consideration is that S3Ms are trained exclusively on raw audio signals, without access to text or explicit semantic labels. This foundational premise is strongly supported by experimental evidence from a recent study \cite{SSL_phonetic}, which demonstrates that S3M representations are substantially more sensitive to phonetic similarities than to semantic ones. 
This observation reinforces our central thesis: the model's advanced capabilities do not stem from a true comprehension of abstract grammar or concepts, but instead \textit{emerge} from learning statistical co-occurrence patterns of speech from vast data. In effect, S3Ms approximate language understanding by constructing \textit{pseudo-linguistic} properties in representations learned from phonetic patterns.

Based on this perspective, we refer to the model's intrinsic \textit{capability} to understand and model speech as the \textit{phonological prior}—a form of knowledge fundamentally based on phonetic patterns. This prior spans both local phonotactic constraints and broader statistical regularities across longer contexts that approximate lexical and syntactic structures. To be precise, in our framework, the phonological prior is the knowledge encoded within the model's pre-trained weights, while the pseudo-linguistic properties are the manifest attributes of the high-level phonetic representations produced by this prior.

\subsection{Self-Supervised Speech Models for Speech Enhancement}
\begin{figure*}[t]
  \centering
  \centerline{\includegraphics[width=1\linewidth]{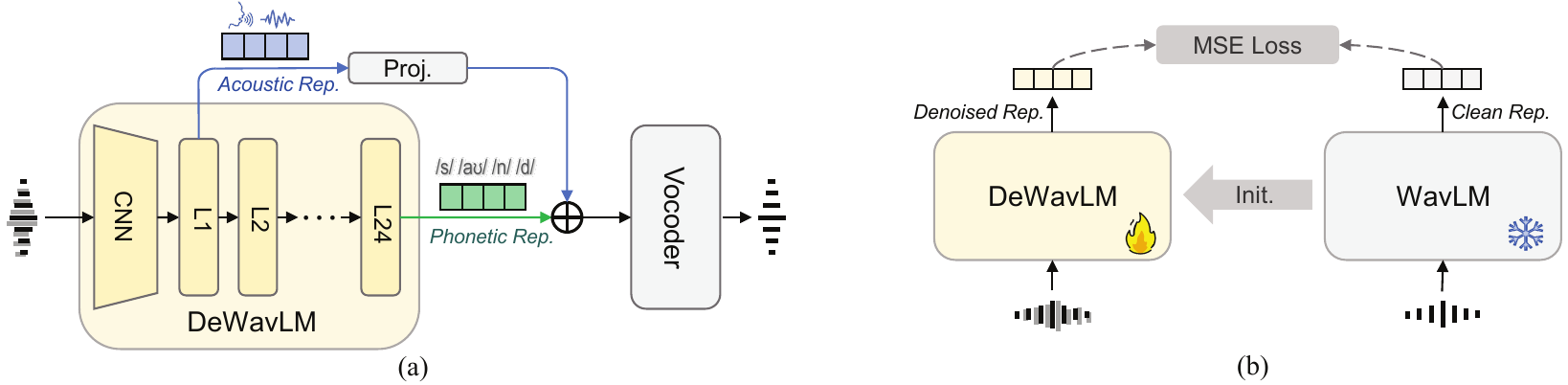}}
  \caption{(a) Overall architecture of the proposed PASE framework. (b) Diagram of denoising representation distillation.}
  \label{fig:framework}
\end{figure*}

The advent of S3Ms has catalyzed the evolution of speech enhancement towards generative paradigms capable of delivering superior perceptual quality. Within this landscape, two major approaches have emerged, differing fundamentally in how they handle S3M representations.

The first paradigm, which we term continuous representation mapping (CRM), operates directly in the continuous domain. It typically employs a vocoder to reconstruct enhanced waveforms from either raw noisy representations \cite{DeVo} or their denoised counterparts produced by a mapping network \cite{Efficient_SSL_SE}. However, we posit that CRM approaches treat these representations as mere sequences of feature vectors, overlooking the contextual structure that underpins their pseudo-linguistic properties. While these approaches outperform traditional spectrogram-based methods \cite{Investigating_SSL_for_SE}, their gains likely stem from the inherent strengths of high-level representations rather than effective modeling of their internal structure.

The second paradigm, termed discrete language modeling (DLM), discretizes S3M representations into token sequences and then attempts to explicitly learn a phonological prior by modeling token distributions autoregressively \cite{Genhancer, GenSE}. While correctly identifying the importance of context, we argue that this approach risks prior contamination: by learning from corrupted representations, the model is prone to generating linguistically inconsistent outputs. 
Some DLM variants, such as SELM \cite{SELM} and LLaSE-G1 \cite{LLaSE-G1}, adopt a parallel prediction paradigm, framing denoising as a masked prediction task in order to encourage the model to capture contextual dependencies when inferring the “masked” (i.e., noisy) tokens. However, we contend this analogy is flawed. Unlike a true hard mask that completely obscures information, noise acts more like a soft mask that merely distorts it. This crucial distinction often allows the model to reconstruct the output from local cues alone, bypassing the need to leverage the very contextual knowledge.
The above limitations in current approaches leave the core challenge of leveraging S3M's long-range contextual knowledge for robust denoising.

\section{Methods}
\subsection{Framework Overview}
In this section, we introduce PASE, a generative SE framework designed to deliver high perceptual quality while effectively mitigating hallucinations. As depicted in Fig.~\ref{fig:framework}~(a), the PASE framework consists of two key components: (1) Denoising WavLM (DeWavLM), created by fine-tuning WavLM through a denoising representation distillation (DRD) strategy, adapting it into a denoising expert that produces enhanced representations given noisy inputs; (2) A vocoder, which reconstructs the enhanced waveform from DeWavLM's dual-stream representations.

This dual-stream design is motivated by the limitations of existing layer-selection strategies for SE. Weighted sum across all layers often overemphasizes the first layer \cite{DeVo}, limiting access to high-level phonetic information. Alternatively, empirically selecting a single intermediate layer (e.g., the 6th) \cite{GenSE, LLaSE-G1}, which mixes acoustic and phonetic cues, yielded suboptimal performance in our preliminary experiments, likely due to insufficient separation of the two types of information.

In contrast, PASE adopts a principled approach. Drawing from established layer-wise analyses \cite{Comparative_layerwise_SSL_analysis}, we select two specialized representations from the DeWavLM module: (1) \textbf{Phonetic representation}: The output of the final transformer layer, which is rich in abstract, context-dependent phonetic content. (2) \textbf{Acoustic representation}: The output of the first transformer layer, which retains the fine-grained acoustic details crucial for preserving speaker identity and prosody.

\subsection{Denoising Representation Distillation}
The DRD strategy, as illustrated in Fig.~\ref{fig:framework}~(b), is a feature-level knowledge distillation scheme. We instantiate two copies of the WavLM model: a frozen teacher and a trainable student, both initialized from the pre-trained weights to inherit the phonological prior. The student model is trained to map a noisy input waveform to a clean representation by minimizing the mean-squared error (MSE) loss against the target representation. This target is generated by the teacher model from the corresponding clean waveform. A comprehensive ablation study (see Appendix~\ref{app:ablation_layers}) reveals that using the final layer outputs from both the student and teacher models yields the best performance. Accordingly, all subsequent experiments adopt this configuration.

A crucial concern during DRD is the risk of catastrophic forgetting, where the new denoising objective may overwrite the model's original phonological prior. To mitigate this, we explore a joint objective combining the distillation loss with the original masked prediction loss \cite{WavLM}, weighted equally. Surprisingly, our further ablation studies reveal that the simpler distillation objective is not only adequate but also yields superior performance.

\subsection{Acoustic-Conditioned Reconstruction}
While DRD excels at producing linguistically accurate representations, these high-level features often lack the fine-grained acoustic details necessary for speaker-consistent and high-fidelity waveform synthesis. To bridge this gap, we employ a dual-stream conditioning mechanism to guide a neural vocoder, ensuring the synthesized speech is both intelligible and faithful to the original speaker's characteristics.

Specifically, we adopt a simple yet effective addition strategy: the acoustic representation is first passed through a linear projection layer to align its feature space with that of the phonetic representation, after which they are aggregated via element-wise summation. While more complex fusion strategies like concatenation, cross-attention, and FiLM \cite{FiLM} can be utilized, our further ablation studies confirm that simple addition is sufficient.

For the vocoder backbone, we adopt Vocos \cite{Vocos}, a cutting-edge neural vocoder offering a strong trade-off between synthesis quality and efficiency. We use the improved variant proposed in WavTokenizer \cite{WavTokenizer}, which integrates an attention module to enhance contextual modeling.
To improve perceptual quality, we employ adversarial training using both the multi-period discriminator (MPD) \cite{HiFiGAN} and the multi-band multi-scale STFT discriminator (MBMSD) \cite{DAC}, with their adversarial losses equally weighted. Following Gesper \cite{Gesper}, our training objective combines reconstruction, adversarial, and feature-matching losses with weights of 15, 2, and 1, respectively.

\section{Experiments}
\subsection{Experimental Setup}
\subsubsection{Datasets}
We construct our training dataset using the large-scale corpora provided by the Interspeech 2025 URGENT Challenge (URGENT2) \cite{URGENT2}. Clean speech is drawn from multiple sources, including the LibriVox subset of the DNS5 Challenge \cite{DNS5}, LibriTTS \cite{LibriTTS}, VCTK \cite{VCTK}, and Common Voice 19.0 \cite{CommonVoice}, totaling approximately 2,000 hours. Noise data comes from DNS5, WHAM! \cite{WHAM}, FSD50K \cite{FSD50K}, and FMA \cite{FMA}. Room impulse responses (RIRs) are taken from openSLR26 and openSLR28 \cite{openSLR}. Training mixtures are generated on the fly. For each example, the clean utterance is convolved with a randomly selected RIR with 50\% probability, then mixed with a randomly chosen noise clip at an SNR uniformly sampled between -5 and 15 dB. The training target is obtained by retaining the first 50 ms of reflections.

We construct a test set with transcripts by leveraging speech and metadata from the \textit{test-clean} split of the LibriTTS corpus, enabling WER evaluation. The noise data is sourced from the validation portion of the URGENT2 corpus to prevent any leakage into the training set. In total, 1,000 test samples are generated using the same procedure as in training. 
For ablation studies on phonetic representation quality, we use the \textit{dev-clean} split of LibriSpeech \cite{LibriSpeech}, which provides aligned phonetic transcripts. To facilitate more comprehensive comparisons, we also include the public DNS1 test set \cite{DNS1}, which includes two subsets: \textit{with-reverb} and \textit{without-reverb}, depending on whether the clean speech contains reverberation. All audio samples are resampled to 16 kHz.

\subsubsection{Baselines}
We compare our PASE model with SOTA SE models, including the discriminative model TF-GridNet \cite{TF-GridNet}, the diffusion-based model StoRM \cite{StoRM}, the flow-matching model FlowSE \cite{FlowSE}, the LM-based model LLaSE-G1 \cite{LLaSE-G1}, and the commercial model Adobe Enhance Speech V2 (AES-V2)\footnote{\url{https://podcast.adobe.com/en/enhance}}. 
TF-GridNet is trained from scratch using the official implementation, while all other models are evaluated using their publicly released checkpoints. Further details of the baseline systems are provided in Appendix~\ref{app:baselines}.

\subsubsection{Evaluation Metrics}
We adopt multiple evaluation metrics from the URGENT2 Challenge to comprehensively assess the performance of our proposed PASE model and the baseline systems. The metrics can be categorized as follows: (1) \textbf{Non-intrusive speech enhancement metrics}: DNSMOS \cite{DNSMOS-P835} and UTMOS \cite{UTMOS}; (2) \textbf{Downstream-task-independent metrics}: Levenshtein phoneme similarity (LPS) \cite{LPS} and SpeechBERTScore (SBS) \cite{SpeechBERTScore}; (3) \textbf{Downstream-task-dependent metrics}: speaker similarity (SpkSim) evaluated using RawNet3 \cite{RawNet3}, and word error rate (WER) computed with OWSM v3.1 \cite{OWSM}. Notably, the third category is particularly effective at capturing speech hallucinations, offering insights into the preservation of speaker characteristics and linguistic content. For all metrics except WER, higher values indicate better performance.

\subsubsection{Implementation Details}
\begin{itemize}
    \item \textbf{DeWavLM Configurations}: The DeWavLM module is initialized from the official WavLM-Large checkpoint\footnote{\url{https://github.com/microsoft/unilm/tree/master/wavlm}}. During DRD, the entire model is fine-tuned. When incorporating the masked prediction loss, we generate pseudo labels by clustering the 9th transformer layer output of the released 2nd-iteration HuBERT-Base model\footnote{\url{https://github.com/facebookresearch/fairseq/tree/main/examples/hubert}}.

    \item \textbf{Vocoder Configurations}: The Vocos backbone comprises a linear layer projecting inputs into a 768-dimensional hidden space, followed by an attention module and 12 ConvNeXt blocks \cite{ConvNeXt}, each with a shared intermediate dimension of 2304. The iSTFT uses an FFT size of 1280 and a hop size of 320.
    
    \item \textbf{Training Details}: The DeWavLM model is trained for 100k steps with a batch size of 4 and a learning rate of 1e-4, while the vocoder is trained for 200k steps with a batch size of 12 and a learning rate of 2e-4. During vocoder training, DeWavLM is kept frozen. All models are optimized using AdamW with a linear warm-up over the first 10\% of steps, followed by cosine decay. During training, all utterances are cropped to 4 seconds. All experiments are conducted on 4 NVIDIA RTX 4090 GPUs.
\end{itemize}

\subsection{Ablation Study}
To validate the effectiveness of our key design choices, we first conduct a series of ablation studies on the simulated LibriTTS test set. For brevity, we report only DNSMOS (OVRL), UTMOS, SpkSim, and WER here. Comprehensive evaluation results across all metrics for the ablation studies are presented in Appendix~\ref{app:full_results}.

\subsubsection{On the Distillation Objective}
\label{sec:ablation_objective}
\begin{table}[t]
\centering
\small
\setlength{\tabcolsep}{5.8mm}
\begin{tabular}{@{}cccc@{}}
\toprule
DRD Objective & PNMI $\uparrow$ & RFS $\uparrow$ & MRS $\uparrow$ \\ \midrule
w/o DRD & 0.69 & 1.00 & 0.68 \\ \midrule
SSL & 0.68 & 0.71 & 0.68 \\
KD & 0.69 & \textbf{0.98} & 0.76 \\
SSL+KD & \textbf{0.70} & 0.93 & \textbf{0.77} \\ \bottomrule
\end{tabular}
\caption{Evaluation of phonological prior preservation. ``w/o DRD" denotes the baseline without fine-tuning. PNMI is calculated on LibriSpeech \textit{dev-clean}, while RFS and MRS are measured on clean speech from our LibriTTS test set.}
\label{tab:phonetic}
\end{table}

\begin{table}[t]
\centering
\small
\setlength{\tabcolsep}{1mm}
\begin{tabular}{@{}ccccc@{}}
\toprule
DRD Objective & DNSMOS $\uparrow$ & UTMOS $\uparrow$ & SpkSim $\uparrow$ & WER (\%) $\downarrow$ \\ \midrule
Noisy & 1.33 & 1.44 & 0.77 & 14.35 \\
Clean & 3.02 & 3.26 & 1.00 & 2.60 \\ \midrule
 w/o DRD & 1.55 & 1.39 & 0.46 & 32.33 \\ \midrule
 SSL & 2.64 & 1.96 & 0.39 & 15.38 \\
 KD & \textbf{3.26} & \textbf{3.42} & \textbf{0.57} & \textbf{7.62} \\
 SSL+KD & 3.07 & 2.95 & 0.52 & 8.78 \\ \bottomrule
\end{tabular}
\caption{Evaluation of denoising performance. ``w/o DRD" denotes the baseline without fine-tuning.}`
\label{tab:objective}
\end{table}

We investigate the impact of the objective function of DRD: (1) KD, which uses only the MSE loss; (2) SSL, which uses only the masked prediction loss; (3) SSL+KD, a joint objective that combines both losses. We assess them on two aspects: \textit{prior preservation} and \textit{denoising performance}. For the former, we evaluate the model's behavior on clean speech using three key metrics:
\begin{itemize}
    \item \textbf{Phone-Normalized Mutual Information (PNMI)}: Measures the mutual information between predicted phone units and reference phone labels, normalized by phone label entropy, as introduced in HuBERT. A higher PNMI indicates better phonetic discriminability.
    \item \textbf{Representation Fidelity Score (RFS)}: Cosine similarity between the model's output and the teacher's output on clean speech. A higher RFS indicates better preservation of desired feature properties after fine-tuning.
    \item \textbf{Masked Reconstruction Score (MRS)}: Cosine similarity between the model’s output on masked input and clean input, calculated over the masked regions. A higher MRS reflects a stronger ability to infer missing content based on the surrounding context, which is acquired through the masked prediction training objective.
\end{itemize}
Specifically, PNMI is computed using 500 k-means clusters, while MRS is calculated by applying mask embeddings to the intermediate CNN outputs at predefined masking indices. For denoising performance, the enhanced representations produced by the fine-tuned DeWavLM are fed to a pre-trained vocoder to synthesize the final waveform, on which we then compute objective enhancement metrics.

The results of our prior preservation analysis, as presented in Table~\ref{tab:phonetic}, demonstrate that all fine-tuning objectives retain PNMI scores comparable to the original model (w/o DRD), indicating that phonetic discriminability is well preserved. 
However, applying only the SSL loss causes a notable drop in RFS, indicating a \textit{representation shift}: the learned features deviate from the original manifold toward a new, incompatible one. While this shifted manifold retains core phonetic properties (as evidenced by the stable PNMI), we argue it offers no discernible advantage and risks overfitting to the limited fine-tuning data, discarding the robustness gained from large-scale pretraining.
Adding the KD loss (SSL+KD) effectively mitigates this shift. Acting as a regularizer, it draws the representations back toward the original manifold, restoring RFS to 0.93. Notably, MRS also improves, suggesting that the KD objective reinforces contextual inference. We hypothesize that DRD functions as a form of prior refinement: by providing phonologically grounded supervision, it enables the model to further hone its contextual reasoning by generalizing to speech in adverse acoustic conditions.
Interestingly, the best performance is achieved by the KD-only objective, which delivers a near-perfect RFS of 0.98 and a strong MRS of 0.76. This suggests that teacher guidance alone provides strong regularization, effectively preventing both knowledge degradation and catastrophic forgetting, while avoiding SSL's drawbacks.

The denoising performance results in Table~\ref{tab:objective} align with our previous findings. While the SSL objective can partially reduce noise, the representation shift leads to incompatibility with the pre-trained vocoder, yielding only modest improvements. The suboptimal performance of the joint objective further supports this concern. In contrast, the KD objective achieves a substantial improvement in denoising performance, highlighting its superiority and effectiveness.

\subsubsection{On the Origins of the Phonological Prior}
\label{sec:ablation_origins}
In this section, we first establish the critical role of the phonological prior. We then design a series of experiments to explore a central question: \textit{what gives rise to this prior}? Since the pre-training process involves two key components: \textit{large-scale data exposure} and a \textit{masked prediction objective}, we structure our investigation around these two factors. 

\begin{table}[t]
\centering
\small
\setlength{\tabcolsep}{1mm}
\begin{tabular}{@{}ccccc@{}}
\toprule
DeWavLM & DNSMOS $\uparrow$ & UTMOS $\uparrow$ & SpkSim $\uparrow$ & WER (\%) $\downarrow$ \\ \midrule
Base     & 3.32   & \textbf{3.67}  & 0.50   & 15.49    \\
Base+    & 3.30   & 3.55  & 0.49   & 13.34    \\ 
Large    & 3.26   & 3.42  & \textbf{0.57}   & \textbf{7.62}     \\ \midrule
Base-FS  & \textbf{3.33} & 3.58 & 0.44 & 36.16 \\
Large-FS & 3.24 & 3.20 & 0.41 & 38.62 \\

\bottomrule
\end{tabular}
\caption{Evaluations of denoising performance for different DeWavLM variants.}
\label{tab:dewavlm}
\end{table}

\textit{The Necessity of a Pre-Existing Prior}:
We compare two DRD configurations for the Large-sized model: one initialized from the pre-trained WavLM (Large in Table~\ref{tab:dewavlm}) and the other from scratch (Large-FS), both fine-tuned with KD loss. As shown, the randomly initialized model fails to achieve comparable denoising performance, especially in terms of WER (38.62\% vs. 7.62\%). This substantial gap highlights that the knowledge inherited from pre-training is the most critical factor for mitigating linguistic hallucinations.

\textit{The Role of Data Scale}: 
We now investigate the role of data scale in shaping the phonological prior. A common assumption is that the prior derives primarily from exposure to large-scale data. However, as shown in Table~\ref{tab:dewavlm}, the Base model—initialized from a checkpoint pre-trained on a relatively modest 960-hour dataset—significantly outperforms its randomly initialized counterpart (Base-FS), achieving a WER of 15.49\% vs. 36.16\%. This suggests that an effective phonological prior can be established even with limited data, indicating that data scale is not the fundamental source of the prior. Furthermore, increasing the pre-training data from 960 hours to 94,000 hours (Base+) yields no notable improvement. This suggests that for a Base-sized architecture, simply scaling up data does not proportionally strengthen the prior. In contrast, increasing model capacity (Large, pre-trained on the same 94k-hour dataset) leads to significant gains, highlighting that the full benefit of large-scale data requires sufficient model capacity.

In summary, our analysis reveals that data scale is a critical \textit{amplifier} but not the foundational \textit{source} of the phonological prior. A strong prior can emerge from modest data, but its full power materializes only through the synergy of high-capacity models and massive, diverse training data.

\begin{table}[t]
\centering
\small
\setlength{\tabcolsep}{6.5mm}
\begin{tabular}{@{}ccc@{}}
\toprule
DeWavLM   & RFS $\uparrow$    & MRS $\uparrow$   \\ \midrule
Base               & 0.93   & 0.55    \\
Large              & \textbf{0.96}  & \textbf{0.76}       \\ \midrule
Base-FS            & 0.84  & 0.32   \\ 
Large-FS           & 0.71  & 0.23    \\ \bottomrule
\end{tabular}
\caption{Evaluation of phonological prior richness for different DeWavLM variants.}
\label{tab:mask}
\end{table}

\textit{The Foundational Role of the Masked Prediction Objective}: 
The evidence so far points towards the pre-training objective—masked prediction—as the potential foundational source. We again take advantage of the RFS and MRS metrics to reveal the crucial relationship between masked prediction and prior strength. 
As shown in Table~\ref{tab:mask}, the randomly initialized models (Base-FS and Large-FS) exhibit relatively high RFS, indicating the distillation successfully teaches them to mimic the teacher's representation properties. However, this mimicry is superficial: these models show extremely low MRS, demonstrating a failure to acquire the ability to perform contextual inference.

In contrast, both pre-trained models (Base and Large) achieve significantly higher RFS and MRS, validating their superior ability to recover masked information beyond simply mapping representation properties. 
Notably, the RFS and MRS gaps between the pre-trained and randomly initialized versions are much greater for the Large-sized model. This suggests its embedded pseudo-linguistic properties and phonological prior are more complex and thus harder to learn from scratch through simple distillation.

Finally, we observe a strong correlation between a model's WER in the denoising task (Table~\ref{tab:dewavlm}) and its MRS in this probing task. This relationship, along with the above cues, suggests that the phonological prior is fundamentally sourced from the contextual modeling ability instilled by the masked prediction objective. Such knowledge cannot be acquired through mere feature-level mimicry and represents the indispensable foundation for robust speech restoration.

\subsubsection{On the Acoustic-Conditioned Schemes}
\label{sec:ablation_condition}
\begin{table}[t]
\centering
\small
\setlength{\tabcolsep}{1mm}
\begin{tabular}{@{}ccccc@{}}
\toprule
Method        & DNSMOS $\uparrow$ & UTMOS $\uparrow$ & SpkSim $\uparrow$ & WER (\%) $\downarrow$ \\ \midrule
w/o condition & 3.26   & 3.42  & 0.57   & 7.62     \\ \midrule
Add      & 3.11   & \textbf{3.09}  & \textbf{0.80}   & 7.50        \\
Cat      & 3.12   & \textbf{3.09}  & \textbf{0.80}   & \textbf{7.49}    \\
CA       & \textbf{3.13}   & 3.08  & 0.79   & 7.78    \\ 
FiLM     & 3.10   & 3.06  & \textbf{0.80}   & 7.58 \\ \bottomrule
\end{tabular}
\caption{Evaluation of denoising performance for different acoustic-conditioned schemes.}
\label{tab:acousitc}
\end{table}

In this section, we validate our dual-stream design by analyzing different acoustic conditioning schemes. As shown in Table~\ref{tab:acousitc}, conditioning on acoustic representation causes a moderate drop in UTMOS (from 3.42 to 3.09) due to residual noise in the shallow-layer features, but significantly boosts SpkSim from 0.57 to 0.80. This confirms that injecting low-level acoustic cues is crucial for suppressing acoustic hallucinations.

Interestingly, the fusion scheme—from simple addition (Add) and concatenation (Cat) to complex cross-attention (CA) or FiLM—has little impact on performance. We attribute this to two factors: (1) the phonetic and acoustic representations are largely orthogonal (see Appendix~\ref{app:orthogonality_analysis}), making addition efficient with minimal information loss; and (2) the vocoder has sufficient capacity to implicitly model the interaction between these disentangled features. Given its strong performance and simplicity, we adopt addition as the default fusion strategy in PASE.

\begin{table*}[t]
\centering
\small
\setlength{\tabcolsep}{1mm}
\begin{tabular}{@{}ccccccccccc@{}}
\toprule
\multirow{2}{*}{Model} & \multirow{2}{*}{Params (M)} & \multirow{2}{*}{MACs (G/s)} & \multicolumn{3}{c}{DNSMOS} & \multirow{2}{*}{UTMOS $\uparrow$} & \multirow{2}{*}{SBS $\uparrow$} & \multirow{2}{*}{LPS $\uparrow$} & \multirow{2}{*}{SpkSim $\uparrow$} & \multirow{2}{*}{WER (\%) $\downarrow$} \\ \cmidrule(lr){4-6}
 &  &  & OVRL $\uparrow$ & SIG $\uparrow$ & BAK $\uparrow$ &  &  &  &  &  \\ \midrule
Noisy & - & - & 1.33 & 1.72 & 1.36 & 1.44 & 0.62 & 0.63 & 0.77 & 14.35 \\
Clean & - & - & 3.02 & 3.42 & 3.76 & 3.26 & 1.00 & 1.00 & 1.00 & 2.60 \\ \midrule
TF-GridNet & 2.77 & 49.63 & 3.04 & 3.34 & \underline{3.96} & 2.62 & \underline{0.85} & \underline{0.90} & \textbf{0.80} & \underline{9.93} \\
StoRM & 55.12 & $\textrm{317.76}\times\textrm{30}$ & 3.07 & 3.38 & \underline{3.96} & 2.55 & 0.68 & 0.65 & \underline{0.63} & 45.94 \\
FlowSE & 350.63 & $\textrm{36.79}\times\textrm{32}$ & 2.38 & 2.91 & 3.22 & 1.74 & 0.71 & 0.70 & 0.57 & 30.13 \\
LLaSE-G1 & 1895.63 & 63.90 & \underline{3.16} & \underline{3.51} & 3.88 & \underline{3.17} & 0.74 & 0.71 & 0.42 & 36.58 \\
AES-V2 & - & - & \textbf{3.35} & \textbf{3.56} & \textbf{4.17} & \textbf{4.09} & 0.79 & 0.85 & 0.60 & 21.32 \\ \midrule
PASE (ours) & 382.14 & 21.42 & 3.12 & 3.48 & 3.88 & 3.09 & \textbf{0.90} & \textbf{0.93} & \textbf{0.80} & \textbf{7.49} \\ \bottomrule
\end{tabular}
\caption{Comparison results on the simulated LibriTTS test set.}
\label{tab:libritts}
\end{table*}

\begin{table*}[t]
\centering
\small
\setlength{\tabcolsep}{1mm}
\begin{tabular}{@{}cccccccccccccccc@{}}
\toprule
\multirow{3}{*}{Model} & \multicolumn{7}{c}{Without Reverb} &  & \multicolumn{7}{c}{With Reverb} \\ \cmidrule(lr){2-8} \cmidrule(l){10-16} 
 & \multicolumn{3}{c}{DNSMOS} & \multirow{2}{*}{UTMOS} & \multirow{2}{*}{SBS} & \multirow{2}{*}{LPS} & \multirow{2}{*}{SpkSim} &  & \multicolumn{3}{c}{DNSMOS} & \multirow{2}{*}{UTMOS} & \multirow{2}{*}{SBS} & \multirow{2}{*}{LPS} & \multirow{2}{*}{SpkSim} \\ \cmidrule(lr){2-4} \cmidrule(lr){10-12}
 & OVRL & SIG & BAK &  &  &  &  &  & OVRL & SIG & BAK &  &  &  &  \\ \midrule
Noisy & 2.48 & 3.39 & 2.62 & 2.36 & 0.80 & 0.90 & 0.94 &  & 1.39 & 1.76 & 1.50 & 1.30 & 0.78 & 0.66 & 0.88 \\
Clean & 3.28 & 3.56 & 4.04 & 4.14 & 1.00 & 1.00 & 1.00 &  & 1.86 & 2.33 & 2.26 & 1.38 & 1.00 & 1.00 & 1.00 \\ \midrule
TF-GridNet & 3.35 & 3.58 & \underline{4.17} & 3.86 & \underline{0.92} & \textbf{0.97} & \textbf{0.94} &  & 2.63 & 3.04 & 3.66 & 1.41 & \underline{0.81} & \underline{0.84} & \textbf{0.84} \\
StoRM & 3.31 & 3.58 & 4.08 & 3.73 & 0.89 & \underline{0.95} & \underline{0.93} &  & 2.63 & 3.03 & 3.82 & 1.48 & 0.47 & 0.27 & 0.32 \\
FlowSE & 3.27 & 3.52 & 4.10 & 3.09 & 0.85 & 0.91 & 0.82 &  & 2.25 & 2.81 & 3.06 & 1.36 & \textbf{0.82} & 0.73 & 0.72 \\
LLaSE-G1 & \textbf{3.42} & \textbf{3.67} & 4.14 & 3.84 & 0.84 & 0.90 & 0.77 &  & \underline{3.35} & \textbf{3.60} & \underline{4.10} & \underline{2.90} & 0.69 & 0.67 & 0.51 \\
AES-V2 & \textbf{3.42} & 3.61 & \textbf{4.20} & \textbf{4.08} & 0.88 & 0.94 & 0.75 &  & \textbf{3.40} & \underline{3.59} & \textbf{4.20} & \textbf{3.71} & 0.70 & 0.79 & 0.67 \\ \midrule
PASE (ours) & \underline{3.39} & \underline{3.63} & 4.15 & \underline{3.95} & \textbf{0.93} & \textbf{0.97} & \textbf{0.94} &  & 2.75 & 3.22 & 3.61 & 1.61 & \textbf{0.82} & \textbf{0.85} & \underline{0.82} \\ \bottomrule
\end{tabular}
\caption{Comparison results on the DNS1 test set.}
\label{tab:dns1}
\end{table*}

\subsection{Comparison with Baselines}
\subsubsection{Results on the LibriTTS Test Set}
We compare PASE against various SOTA baselines, with results on our simulated test set summarized in Table~\ref{tab:libritts}. The key findings are as follows:
(1) The discriminative TF-GridNet achieves a low WER of 9.93\% and a high SpkSim of 0.80, confirming that discriminative models are less prone to hallucinations. However, this comes at the cost of perceptual quality, as evidenced by its significantly lower UTMOS score of 2.62. 
(2) The LM-based LLaSE-G1 suffers from severe hallucinations, with a markedly high WER of 36.58\% and a poor SpkSim of 0.42, supporting our hypothesis that modeling speech structure from corrupted inputs risks hallucinations.
(3) Other generative models, including the diffusion-based StoRM and the flow-matching-based FlowSE, exhibit similar hallucination problems despite their significantly higher computational complexity.
(4) The commercial AES-V2 delivers the best perceptual quality, with the highest UTMOS of 4.09. However, this comes at the expense of content integrity, evidenced by a high WER of 21.32\%, highlighting the challenge of achieving both perceptual quality and linguistic accuracy.
(5) In contrast, our proposed PASE demonstrates well-balanced performance across all aspects. It maintains high perceptual quality while achieving the lowest WER and the highest SpkSim, LPS, and SBS scores, validating its strong fidelity across acoustic, phonetic, and semantic dimensions—all with the lowest computational cost among the compared methods.

\subsubsection{Results on the DNS1 Test set}
For a more comprehensive evaluation, we assess performance on the DNS1 test set, with results shown in Table~\ref{tab:dns1}. Since DNS1 lacks transcripts, we analyze linguistic integrity using SBS and LPS metrics. On the \textit{without-reverb} subset, performance trends largely align with the previous findings, with PASE ranking first or second across nearly all metrics. However, the gap between PASE and TF-GridNet narrows, likely due to the higher SNR in DNS1, where speech is largely intelligible and the benefit of a generative prior is less pronounced. In contrast, our simulated test set includes extremely low-SNR cases, where a phonological prior is crucial for recovering missing content and avoiding hallucinations.

The \textit{with-reverb} subset presents a more challenging scenario and reveals some distinctions. Models such as LLaSE-G1 and AES-V2 achieve significantly higher scores on perceptual quality metrics. However, this comes at the cost of pronounced degradation in speaker fidelity (SpkSim) and linguistic integrity (LPS and SBS). These results suggest that their approaches produce perceptually ``clean" yet structurally distorted speech, leading to severe hallucinations. 
In contrast, PASE excels consistently in linguistic integrity, achieving the highest LPS (0.85) and SBS (0.82), while maintaining a strong SpkSim (0.82). Regarding the relatively low UTMOS score, we attribute this to two main factors: (1) a potential mismatch in reverberation intensity between the training data and the DNS1 test set, and (2) a potential bias of UTMOS towards ``dry" speech, as evidenced by the substantial score drop from clean speech in the \textit{without-reverb} to the \textit{with-reverb} subset. Notably, TF-GridNet also receives a relatively low UTMOS score. Considering its well-established dereverberation capability, the comparable UTMOS of PASE should not be interpreted as an inability to handle reverberation effectively. 
Representative audio examples covering both test sets and highlighting differences in perceptual quality, speaker fidelity, and linguistic integrity are included in Appendix~\ref{app:audio_examples}.

\section{Conclusion}
In this work, we propose PASE, a novel framework for low-hallucination generative speech enhancement. We begin by categorizing hallucinations into linguistic and acoustic types, and introduce corresponding targeted strategies. To mitigate linguistic hallucination, PASE leverages the strong phonological prior encoded in a pre-trained WavLM via denoising representation distillation, which anchors the enhancement process to the original content and preserves linguistic integrity even under severe noise and reverberation. To combat acoustic hallucination, a dual-stream design conditions reconstruction on both phonetic content and acoustic cues, enabling high-quality synthesis while retaining speaker characteristics. This dual-pronged approach makes PASE a more reliable and faithful solution for real-world speech enhancement, bridging the gap between perceptual quality and content accuracy.

\section{Acknowledgments}
This work was supported by the National Natural Science
Foundation of China (Grant No. 12274221) and the Yangtze River Delta Science and Technology Innovation Community Joint Research Project (Grant No. 2024CSJGG1103).

\bibliography{ref}

@inproceedings{SELM,
  title={SELM: Speech enhancement using discrete tokens and language models},
  author={Wang, Ziqian and Zhu, Xinfa and Zhang, Zihan and Lv, YuanJun and Jiang, Ning and Zhao, Guoqing and Xie, Lei},
  booktitle={ICASSP 2024-2024 IEEE International Conference on Acoustics, Speech and Signal Processing (ICASSP)},
  pages={11561--11565},
  year={2024},
  organization={IEEE}
}

@article{FlowSE,
  title={FlowSE: Efficient and High-Quality Speech Enhancement via Flow Matching},
  author={Wang, Ziqian and Liu, Zikai and Zhu, Xinfa and Zhu, Yike and Liu, Mingshuai and Chen, Jun and Xiao, Longshuai and Weng, Chao and Xie, Lei},
  journal={arXiv preprint arXiv:2505.19476},
  year={2025}
}

@article{StoRM,
  title={StoRM: A diffusion-based stochastic regeneration model for speech enhancement and dereverberation},
  author={Lemercier, Jean-Marie and Richter, Julius and Welker, Simon and Gerkmann, Timo},
  journal={IEEE/ACM Transactions on Audio, Speech, and Language Processing},
  volume={31},
  pages={2724--2737},
  year={2023},
  publisher={IEEE}
}

@inproceedings{CDiffSE,
  title={Conditional diffusion probabilistic model for speech enhancement},
  author={Lu, Yen-Ju and Wang, Zhong-Qiu and Watanabe, Shinji and Richard, Alexander and Yu, Cheng and Tsao, Yu},
  booktitle={ICASSP 2022-2022 IEEE International Conference on Acoustics, Speech and Signal Processing (ICASSP)},
  pages={7402--7406},
  year={2022},
  organization={Ieee}
}

@article{TF-GridNet,
  title={{TF-GridNet}: Integrating full-and sub-band modeling for speech separation},
  author={Wang, Zhong-Qiu and Cornell, Samuele and Choi, Shukjae and Lee, Younglo and Kim, Byeong-Yeol and Watanabe, Shinji},
  journal={IEEE/ACM Transactions on Audio, Speech, and Language Processing},
  volume={31},
  pages={3221--3236},
  year={2023},
  publisher={IEEE}
}

@inproceedings{Genhancer,
  title     = {Genhancer: High-Fidelity Speech Enhancement via Generative Modeling on Discrete Codec Tokens},
  author    = {Haici Yang and Jiaqi Su and Minje Kim and Zeyu Jin},
  year      = {2024},
  booktitle = {Interspeech 2024},
  pages     = {1170--1174},
  doi       = {10.21437/Interspeech.2024-590},
  issn      = {2958-1796},
}

@inproceedings{SSL_phonetic,
  title     = {Self-Supervised Speech Representations are More Phonetic than Semantic},
  author    = {Kwanghee Choi and Ankita Pasad and Tomohiko Nakamura and Satoru Fukayama and Karen Livescu and Shinji Watanabe},
  year      = {2024},
  booktitle = {Interspeech 2024},
  pages     = {4578--4582},
  doi       = {10.21437/Interspeech.2024-1157},
  issn      = {2958-1796},
}

@article{HuBERT,
  title={{HuBERT}: Self-supervised speech representation learning by masked prediction of hidden units},
  author={Hsu, Wei-Ning and Bolte, Benjamin and Tsai, Yao-Hung Hubert and Lakhotia, Kushal and Salakhutdinov, Ruslan and Mohamed, Abdelrahman},
  journal={IEEE/ACM transactions on audio, speech, and language processing},
  volume={29},
  pages={3451--3460},
  year={2021},
  publisher={IEEE}
}

@article{WavLM,
  title={{WavLM}: Large-scale self-supervised pre-training for full stack speech processing},
  author={Chen, Sanyuan and Wang, Chengyi and Chen, Zhengyang and Wu, Yu and Liu, Shujie and Chen, Zhuo and Li, Jinyu and Kanda, Naoyuki and Yoshioka, Takuya and Xiao, Xiong and others},
  journal={IEEE Journal of Selected Topics in Signal Processing},
  volume={16},
  number={6},
  pages={1505--1518},
  year={2022},
  publisher={IEEE}
}

@article{DAC,
  title={High-fidelity audio compression with improved {RVQGAN}},
  author={Kumar, Rithesh and Seetharaman, Prem and Luebs, Alejandro and Kumar, Ishaan and Kumar, Kundan},
  journal={Advances in Neural Information Processing Systems},
  volume={36},
  pages={27980--27993},
  year={2023}
}

@article{LLaSE-G1,
  title={{LLaSE-G1}: Incentivizing generalization capability for llama-based speech enhancement},
  author={Kang, Boyi and Zhu, Xinfa and Zhang, Zihan and Ye, Zhen and Liu, Mingshuai and Wang, Ziqian and Zhu, Yike and Ma, Guobin and Chen, Jun and Xiao, Longshuai and others},
  journal={arXiv preprint arXiv:2503.00493},
  year={2025}
}

@article{GenSE,
  title={{GenSE}: Generative Speech Enhancement via Language Models using Hierarchical Modeling},
  author={Yao, Jixun and Liu, Hexin and Chen, Chen and Hu, Yuchen and Chng, EngSiong and Xie, Lei},
  journal={arXiv preprint arXiv:2502.02942},
  year={2025}
}

@inproceedings{DNSMOS,
 author = {Chandan K. A. Reddy and
Vishak Gopal and
Ross Cutler},
 bibsource = {dblp computer science bibliography, https://dblp.org},
 booktitle = {{IEEE} International Conference on Acoustics, Speech and Signal Processing,
{ICASSP} 2021, Toronto, ON, Canada, June 6-11, 2021},
 doi = {10.1109/ICASSP39728.2021.9414878},
 pages = {6493--6497},
 publisher = {{IEEE}},
 title = {{DNSMOS}: {A} Non-Intrusive Perceptual Objective Speech Quality Metric
to Evaluate Noise Suppressors},
 year = {2021}
}

@inproceedings{DNSMOS-P835,
 author = {Chandan K. A. Reddy and
Vishak Gopal and
Ross Cutler},
 bibsource = {dblp computer science bibliography, https://dblp.org},
 booktitle = {{IEEE} International Conference on Acoustics, Speech and Signal Processing,
{ICASSP} 2022, Virtual and Singapore, 23-27 May 2022},
 doi = {10.1109/ICASSP43922.2022.9746108},
 pages = {886--890},
 publisher = {{IEEE}},
 title = {{DNSMOS} {P.835:} {A} Non-Intrusive Perceptual Objective Speech Quality
Metric to Evaluate Noise Suppressors},
 year = {2022}
}

@inproceedings{UTMOS,
 author = {Takaaki Saeki and
Detai Xin and
Wataru Nakata and
Tomoki Koriyama and
Shinnosuke Takamichi and
Hiroshi Saruwatari},
 bibsource = {dblp computer science bibliography, https://dblp.org},
 booktitle = {Interspeech 2022},
 pages = {4521--4525},
 publisher = {{ISCA}},
 title = {{UTMOS:} UTokyo-SaruLab System for VoiceMOS Challenge 2022},
 year = {2022}
}

@article{URGENT2,
  title={Interspeech 2025 {URGENT} speech enhancement challenge},
  author={Saijo, Kohei and Zhang, Wangyou and Cornell, Samuele and Scheibler, Robin and Li, Chenda and Ni, Zhaoheng and Kumar, Anurag and Sach, Marvin and Fu, Yihui and Wang, Wei and others},
  journal={arXiv preprint arXiv:2505.23212},
  year={2025}
}

@inproceedings{Eval_generative,
  title={Evaluation metrics for generative speech enhancement methods: Issues and perspectives},
  author={Pirklbauer, Jan and Sach, Marvin and Fluyt, Kristoff and Tirry, Wouter and Wardah, Wafaa and Moeller, Sebastian and Fingscheidt, Tim},
  booktitle={Speech Communication; 15th ITG Conference},
  pages={265--269},
  year={2023},
  organization={VDE}
}

@article{Vocos,
  title={Vocos: Closing the gap between time-domain and fourier-based neural vocoders for high-quality audio synthesis},
  author={Siuzdak, Hubert},
  journal={arXiv preprint arXiv:2306.00814},
  year={2023}
}

@inproceedings{SSL_AR,
  title     = {An Unsupervised Autoregressive Model for Speech Representation Learning},
  author    = {Yu-An Chung and Wei-Ning Hsu and Hao Tang and James Glass},
  year      = {2019},
  booktitle = {Interspeech 2019},
  pages     = {146--150},
  doi       = {10.21437/Interspeech.2019-1473},
  issn      = {2958-1796},
}

@inproceedings{Phonetic_analysis,
  title     = {{Phonetic Analysis of Self-supervised Representations of English Speech}},
  author    = {{Dan Wells and Hao Tang and Korin Richmond}},
  year      = {{2022}},
  booktitle = {{Interspeech 2022}},
  pages     = {{3583--3587}},
  doi       = {{10.21437/Interspeech.2022-10884}},
  issn      = {{2958-1796}},
}

@inproceedings{Phonology_analysis,
  title     = {Probing Self-supervised Speech Models for Phonetic and Phonemic Information: A Case Study in Aspiration},
  author    = {Kinan Martin and Jon Gauthier and Canaan Breiss and Roger Levy},
  year      = {2023},
  booktitle = {Interspeech 2023},
  pages     = {251--255},
  doi       = {10.21437/Interspeech.2023-2359},
  issn      = {2958-1796},
}

@inproceedings{Syntax_analysis,
  title     = {Wave to Syntax: Probing spoken language models for syntax},
  author    = {Gaofei Shen and Afra Alishahi and Arianna Bisazza and Grzegorz Chrupała},
  year      = {2023},
  booktitle = {Interspeech 2023},
  pages     = {1259--1263},
  doi       = {10.21437/Interspeech.2023-679},
  issn      = {2958-1796},
}

@inproceedings{Semantic_analysis,
  title     = {SpeechGLUE: How Well Can Self-Supervised Speech Models Capture Linguistic Knowledge?},
  author    = {Takanori Ashihara and Takafumi Moriya and Kohei Matsuura and Tomohiro Tanaka and Yusuke Ijima and Taichi Asami and Marc Delcroix and Yukinori Honma},
  year      = {2023},
  booktitle = {Interspeech 2023},
  pages     = {2888--2892},
  doi       = {10.21437/Interspeech.2023-1823},
  issn      = {2958-1796},
}

@inproceedings{Layerwise_SSL_analysis,
  title={Layer-wise analysis of a self-supervised speech representation model},
  author={Pasad, Ankita and Chou, Ju-Chieh and Livescu, Karen},
  booktitle={2021 IEEE Automatic Speech Recognition and Understanding Workshop (ASRU)},
  pages={914--921},
  year={2021},
  organization={IEEE}
}

@inproceedings{Comparative_layerwise_SSL_analysis,
  title={Comparative layer-wise analysis of self-supervised speech models},
  author={Pasad, Ankita and Shi, Bowen and Livescu, Karen},
  booktitle={ICASSP 2023-2023 IEEE International Conference on Acoustics, Speech and Signal Processing (ICASSP)},
  pages={1--5},
  year={2023},
  organization={IEEE}
}

@inproceedings{Investigating_SSL_for_SE,
  title={Investigating self-supervised learning for speech enhancement and separation},
  author={Huang, Zili and Watanabe, Shinji and Yang, Shu-wen and Garc{\'\i}a, Paola and Khudanpur, Sanjeev},
  booktitle={ICASSP 2022-2022 IEEE International Conference on Acoustics, Speech and Signal Processing (ICASSP)},
  pages={6837--6841},
  year={2022},
  organization={IEEE}
}

@article{Efficient_SSL_SE,
  title={Efficient Speech Enhancement via Embeddings from Pre-trained Generative Audioencoders},
  author={Sun, Xingwei and Dinkel, Heinrich and Niu, Yadong and Wang, Linzhang and Zhang, Junbo and Luan, Jian},
  journal={arXiv preprint arXiv:2506.11514},
  year={2025}
}

@inproceedings{DeVo,
  title={Self-supervised learning for speech enhancement through synthesis},
  author={Irvin, Bryce and Stamenovic, Marko and Kegler, Mikolaj and Yang, Li-Chia},
  booktitle={ICASSP 2023-2023 IEEE International Conference on Acoustics, Speech and Signal Processing (ICASSP)},
  pages={1--5},
  year={2023},
  organization={IEEE}
}

@inproceedings{FiLM,
  title={{FiLM}: Visual reasoning with a general conditioning layer},
  author={Perez, Ethan and Strub, Florian and De Vries, Harm and Dumoulin, Vincent and Courville, Aaron},
  booktitle={Proceedings of the AAAI conference on artificial intelligence},
  volume={32},
  number={1},
  year={2018}
}

@article{Transformer,
  title={Attention is all you need},
  author={Vaswani, Ashish and Shazeer, Noam and Parmar, Niki and Uszkoreit, Jakob and Jones, Llion and Gomez, Aidan N and Kaiser, {\L}ukasz and Polosukhin, Illia},
  journal={Advances in neural information processing systems},
  volume={30},
  year={2017}
}

@article{DNS5,
  title={{ICASSP} 2023 deep noise suppression challenge},
  author={Dubey, Harishchandra and Aazami, Ashkan and Gopal, Vishak and Naderi, Babak and Braun, Sebastian and Cutler, Ross and Ju, Alex and Zohourian, Mehdi and Tang, Min and Golestaneh, Mehrsa and others},
  journal={IEEE Open Journal of Signal Processing},
  volume={5},
  pages={725--737},
  year={2024},
  publisher={IEEE}
}

@inproceedings{VCTK,
  title={{The Voice Bank} corpus: Design, collection and data analysis of a large regional accent speech database},
  author={Veaux, Christophe and Yamagishi, Junichi and King, Simon},
  booktitle={2013 international conference oriental COCOSDA held jointly with 2013 conference on Asian spoken language research and evaluation (O-COCOSDA/CASLRE)},
  pages={1--4},
  year={2013},
  organization={IEEE}
}

@inproceedings{LibriTTS,
  title     = {{LibriTTS}: A Corpus Derived from {LibriSpeech} for Text-to-Speech},
  author    = {Heiga Zen and Viet Dang and Rob Clark and Yu Zhang and Ron J. Weiss and Ye Jia and Zhifeng Chen and Yonghui Wu},
  year      = {2019},
  booktitle = {Interspeech 2019},
  pages     = {1526--1530},
  doi       = {10.21437/Interspeech.2019-2441},
  issn      = {2958-1796},
}

@inproceedings{LibriSpeech,
  title={{LibriSpeech}: an {ASR} corpus based on public domain audio books},
  author={Panayotov, Vassil and Chen, Guoguo and Povey, Daniel and Khudanpur, Sanjeev},
  booktitle={2015 IEEE international conference on acoustics, speech and signal processing (ICASSP)},
  pages={5206--5210},
  year={2015},
  organization={IEEE}
}

@inproceedings{CommonVoice,
  title={{Common Voice}: A Massively-Multilingual Speech Corpus},
  author={Ardila, Rosana and Branson, Megan and Davis, Kelly and Kohler, Michael and Meyer, Josh and Henretty, Michael and Morais, Reuben and Saunders, Lindsay and Tyers, Francis and Weber, Gregor},
  booktitle={Proceedings of the Twelfth Language Resources and Evaluation Conference},
  pages={4218--4222},
  year={2020}
}

@inproceedings{WHAM,
  title     = {{WHAM!}: Extending Speech Separation to Noisy Environments},
  author    = {Gordon Wichern and Joe Antognini and Michael Flynn and Licheng Richard Zhu and Emmett McQuinn and Dwight Crow and Ethan Manilow and Jonathan Le Roux},
  year      = {2019},
  booktitle = {Interspeech 2019},
  pages     = {1368--1372},
  doi       = {10.21437/Interspeech.2019-2821},
  issn      = {2958-1796},
}

@article{FSD50K,
  title={{FSD50K}: an open dataset of human-labeled sound events},
  author={Fonseca, Eduardo and Favory, Xavier and Pons, Jordi and Font, Frederic and Serra, Xavier},
  journal={IEEE/ACM Transactions on Audio, Speech, and Language Processing},
  volume={30},
  pages={829--852},
  year={2021},
  publisher={IEEE}
}

@article{FMA,
  title={{FMA}: A dataset for music analysis},
  author={Defferrard, Micha{\"e}l and Benzi, Kirell and Vandergheynst, Pierre and Bresson, Xavier},
  journal={arXiv preprint arXiv:1612.01840},
  year={2016}
}

@inproceedings{openSLR,
  title={A study on data augmentation of reverberant speech for robust speech recognition},
  author={Ko, Tom and Peddinti, Vijayaditya and Povey, Daniel and Seltzer, Michael L and Khudanpur, Sanjeev},
  booktitle={2017 IEEE international conference on acoustics, speech and signal processing (ICASSP)},
  pages={5220--5224},
  year={2017},
  organization={IEEE}
}

@inproceedings{DNS1,
  title     = {The INTERSPEECH 2020 Deep Noise Suppression Challenge: Datasets, Subjective Testing Framework, and Challenge Results},
  author    = {Chandan K.A. Reddy and Vishak Gopal and Ross Cutler and Ebrahim Beyrami and Roger Cheng and Harishchandra Dubey and Sergiy Matusevych and Robert Aichner and Ashkan Aazami and Sebastian Braun and Puneet Rana and Sriram Srinivasan and Johannes Gehrke},
  year      = {2020},
  booktitle = {Interspeech 2020},
  pages     = {2492--2496},
  doi       = {10.21437/Interspeech.2020-3038},
  issn      = {2958-1796},
}

@inproceedings{WavTokenizer,
  title={{WavTokenizer}: an Efficient Acoustic Discrete Codec Tokenizer for Audio Language Modeling},
  author={Ji, Shengpeng and Jiang, Ziyue and Wang, Wen and Chen, Yifu and Fang, Minghui and Zuo, Jialong and Yang, Qian and Cheng, Xize and Wang, Zehan and Li, Ruiqi and others},
  booktitle={The Thirteenth International Conference on Learning Representations},
  year={2024}
}

@inproceedings{ConvNeXt,
  title={A convnet for the 2020s},
  author={Liu, Zhuang and Mao, Hanzi and Wu, Chao-Yuan and Feichtenhofer, Christoph and Darrell, Trevor and Xie, Saining},
  booktitle={Proceedings of the IEEE/CVF conference on computer vision and pattern recognition},
  pages={11976--11986},
  year={2022}
}

@article{HiFiGAN,
  title={{HiFi-GAN}: Generative adversarial networks for efficient and high fidelity speech synthesis},
  author={Kong, Jungil and Kim, Jaehyeon and Bae, Jaekyoung},
  journal={Advances in neural information processing systems},
  volume={33},
  pages={17022--17033},
  year={2020}
}

@inproceedings{Gesper,
  title={Gesper: A unified framework for general speech restoration},
  author={Chen, Jun and Shi, Yupeng and Liu, Wenzhe and Rao, Wei and He, Shulin and Li, Andong and Wang, Yannan and Wu, Zhiyong and Shang, Shidong and Zheng, Chengshi},
  booktitle={ICASSP 2023-2023 IEEE International Conference on Acoustics, Speech and Signal Processing (ICASSP)},
  pages={1--2},
  year={2023},
  organization={IEEE}
}

@inproceedings{LPS,
 author = {Pirklbauer, Jan and Sach, Marvin and Fluyt, Kristoff and Tirry, Wouter and Wardah, Wafaa and Moeller, Sebastian and Fingscheidt, Tim},
 booktitle = {Speech Communication; 15th ITG Conference},
 organization = {VDE},
 pages = {265--269},
 title = {Evaluation metrics for generative speech enhancement methods: Issues and perspectives},
 year = {2023}
}

@inproceedings{SpeechBERTScore,
  title     = {SpeechBERTScore: Reference-Aware Automatic Evaluation of Speech Generation Leveraging NLP Evaluation Metrics},
  author    = {Takaaki Saeki and Soumi Maiti and Shinnosuke Takamichi and Shinji Watanabe and Hiroshi Saruwatari},
  year      = {2024},
  booktitle = {Interspeech 2024},
  pages     = {4943--4947},
  doi       = {10.21437/Interspeech.2024-1508},
  issn      = {2958-1796},
}

@inproceedings{RawNet3,
  title     = {{Pushing the limits of raw waveform speaker recognition}},
  author    = {Jee-weon Jung and Youjin Kim and Hee-Soo Heo and Bong-Jin Lee and Youngki Kwon and Joon Son Chung},
  year      = {{2022}},
  booktitle = {{Interspeech 2022}},
  pages     = {{2228--2232}},
  doi       = {{10.21437/Interspeech.2022-126}},
  issn      = {{2958-1796}},
}

@inproceedings{OWSM,
  title     = {{OWSM v3.1}: Better and Faster Open Whisper-Style Speech Models based on E-Branchformer},
  author    = {Yifan Peng and Jinchuan Tian and William Chen and Siddhant Arora and Brian Yan and Yui Sudo and Muhammad Shakeel and Kwanghee Choi and Jiatong Shi and Xuankai Chang and Jee-weon Jung and Shinji Watanabe},
  year      = {2024},
  booktitle = {Interspeech 2024},
  pages     = {352--356},
  doi       = {10.21437/Interspeech.2024-1194},
  issn      = {2958-1796},
}
\appendix
\section{Full Results of Ablation Experiments}
\label{app:full_results}
\subsection{On the Distillation Layers}
\label{app:ablation_layers}
\begin{table*}[!htbp]
\centering
\small
\begin{tabular}{@{}cccccccccc@{}}
\toprule
\multirow{2}{*}{Student Layer} & \multirow{2}{*}{Teacher Layer} & \multicolumn{3}{c}{DNSMOS} & \multirow{2}{*}{UTMOS $\uparrow$} & \multirow{2}{*}{SBS $\uparrow$} & \multirow{2}{*}{LPS $\uparrow$} & \multirow{2}{*}{SpkSim $\uparrow$} & \multirow{2}{*}{WER (\%) $\downarrow$} \\ \cmidrule(lr){3-5}
 &  & OVRL $\uparrow$ & SIG $\uparrow$ & BAK $\uparrow$ &  &  &  &  &  \\ \midrule
L24 & L6 & 3.24 & 3.56 & 3.99 & 3.24 & 0.79 & 0.86 & \textbf{0.74} & 16.96 \\
L24 & L12 & 3.23 & 3.54 & 4.00 & 3.29 & 0.86 & 0.89 & 0.67 & 14.62 \\
L24 & L24 & \textbf{3.26} & \textbf{3.57} & \textbf{4.03} & \textbf{3.42} & \textbf{0.88} & \textbf{0.93} & 0.57 & \textbf{7.62} \\ \midrule
L6 & L6 & 3.23 & 3.56 & 3.98 & 3.18 & 0.83 & 0.84 & \textbf{0.74} & 17.40 \\
L12 & L12 & 3.24 & 3.55 & 4.00 & 3.30 & 0.86 & 0.88 & 0.68 & 14.87 \\ \bottomrule
\end{tabular}
\caption{Evaluation of denoising performance for different configurations of student output layers and teacher target layers.}
\label{tab:layer_full}
\end{table*}

We investigate the optimal distillation layers for denoising representation distillation (DRD) by analyzing both the student and teacher output layers. To disentangle their interaction, we design two experimental settings: (1) \textit{heterogeneous-layer} distillation, where a 24-layer student is fixed and the teacher layer is varied, allowing us to assess how different target representations affect training;
(2) \textit{homogeneous-layer} distillation, where the student learns from the teacher layer at the same depth, isolating the impact of representation mismatch caused by cross-layer supervision.
To evaluate denoising performance, the enhanced representations produced by the student are fed to a vocoder trained specifically on the corresponding teacher layer, ensuring alignment between representation and synthesis.

The results in Table~\ref{tab:layer_full} highlight two key findings. 
First, in the heterogeneous setup (top three rows), performance in linguistic accuracy (reflected by SBS, LPS, and WER) improves consistently with teacher depth. This confirms that deeper teacher layers, which encode richer phonetic information, provide more effective supervision for linguistically faithful denoising. While SpkSim tends to decline with increasing layer depth, we focus solely on linguistic integrity in this analysis, as speaker similarity is addressed separately during the acoustic-conditioned reconstruction stage.

However, the heterogeneous setup introduces a potential confounder: representation mismatch. To isolate this effect, we compare results across both setups. Notably, the L24$\rightarrow$L6 (heterogeneous) configuration shows only marginal improvements interms of WER over the L6$\rightarrow$L6 (homogeneous) one, despite the student’s substantially higher capacity. This suggests that the representational gap between student and teacher leads to ineffective supervision, resulting in a performance bottleneck.
Based on the above analysis, our final PASE model adopts the L24$\rightarrow$L24 strategy to fully leverage both the student’s model capacity and the teacher’s phonetic richness.

\subsection{On the Distillation Objective}
\label{app:ablation_objectives}
\begin{table*}[!htbp]
\centering
\small
\begin{tabular}{@{}ccccccccc@{}}
\toprule
\multirow{2}{*}{DRD Objective} & \multicolumn{3}{c}{DNSMOS} & \multirow{2}{*}{UTMOS $\uparrow$} & \multirow{2}{*}{SBS $\uparrow$} & \multirow{2}{*}{LPS $\uparrow$} & \multirow{2}{*}{SpkSim $\uparrow$} & \multirow{2}{*}{WER (\%) $\downarrow$} \\ \cmidrule(lr){2-4}
 & OVRL $\uparrow$ & SIG $\uparrow$ & BAK $\uparrow$ &  &  &  &  &  \\ \midrule
Noisy & 1.33 & 1.72 & 1.36 & 1.44 & 0.62 & 0.63 & 0.77 & 14.35 \\
Clean & 3.02 & 3.42 & 3.76 & 3.26 & 1.00 & 1.00 & 1.00 & 2.60 \\ \midrule
w/o DRD & 1.55 & 2.15 & 1.62 & 1.39 & 0.60 & 0.44 & 0.46 & 32.33 \\ \midrule
SSL & 2.64 & 3.31 & 3.16 & 1.96 & 0.75 & 0.84 & 0.39 & 15.38 \\
KD & \textbf{3.26} & \textbf{3.57} & \textbf{4.03} & \textbf{3.42} & \textbf{0.88} & \textbf{0.93} & \textbf{0.57} & \textbf{7.62} \\
SSL+KD & 3.07 & 3.42 & 3.95 & 2.95 & 0.86 & 0.92 & 0.52 & 8.78 \\ \bottomrule
\end{tabular}
\caption{Evaluation of denoising performance for different DRD objectives. ``w/o DRD" denotes the baseline without fine-tuning.}
\label{tab:objective_full}
\end{table*}

As the ablation experiments and analysis of the distillation objective are already covered in the main paper, we provide the full metric results here, as presented in Table~\ref{tab:objective_full}. These results consistently reinforce our previous findings.

\subsection{On the Origins of the Phonological Prior}
\begin{table*}[!htbp]
\centering
\small
\begin{tabular}{@{}ccccccccc@{}}
\toprule
\multirow{2}{*}{DeWavLM} & \multicolumn{3}{c}{DNSMOS} & \multirow{2}{*}{UTMOS $\uparrow$} & \multirow{2}{*}{SBS $\uparrow$} & \multirow{2}{*}{LPS $\uparrow$} & \multirow{2}{*}{SpkSim $\uparrow$} & \multirow{2}{*}{WER (\%) $\downarrow$} \\ \cmidrule(lr){2-4}
 & OVRL $\uparrow$ & SIG $\uparrow$ & BAK $\uparrow$ &  &  &  &  &  \\ \midrule
Base & 3.32 & \textbf{3.62} & 4.04 & \textbf{3.67} & 0.84 & 0.88 & 0.50 & 15.49 \\
Base+ & 3.30 & 3.60 & 4.03 & 3.55 & 0.85 & 0.90 & 0.49 & 13.34 \\
Large & 3.26 & 3.57 & 4.03 & 3.42 & \textbf{0.88} & \textbf{0.93} & \textbf{0.57} & \textbf{7.62} \\ \midrule
Base-FS & \textbf{3.33} & \textbf{3.62} & \textbf{4.05} & 3.58 & 0.76 & 0.73 & 0.44 & 36.16 \\
Large-FS & 3.24 & 3.54 & 4.01 & 3.20 & 0.74 & 0.71 & 0.41 & 38.62 \\ \bottomrule
\end{tabular}
\caption{Evaluation of denoising performance for different DeWavLM variants. ``FS" denotes the baseline initialized from scratch.}
\label{tab:origins}
\end{table*}

As the ablation experiments and analysis on the origins of the phonological prior have already been presented in the main paper, we report the full metric results here in Table~\ref{tab:origins}. These results also provide consistent support for our previous findings.

\subsection{On the Acoustic-Conditioned Schemes}

\begin{table*}[!htbp]
\centering
\small
\begin{tabular}{@{}ccccccccc@{}}
\toprule
\multirow{2}{*}{Method} & \multicolumn{3}{c}{DNSMOS} & \multirow{2}{*}{UTMOS $\uparrow$} & \multirow{2}{*}{SBS $\uparrow$} & \multirow{2}{*}{LPS $\uparrow$} & \multirow{2}{*}{SpkSim $\uparrow$} & \multirow{2}{*}{WER (\%) $\downarrow$} \\ \cmidrule(lr){2-4}
 & OVRL $\uparrow$ & SIG $\uparrow$ & BAK $\uparrow$ &  &  &  &  &  \\ \midrule
w/o condition & \textbf{3.26} & \textbf{3.57} & \textbf{4.03} & \textbf{3.42} & 0.88 & \textbf{0.93} & 0.57 & 7.62 \\ \midrule
Add & 3.11 & 3.48 & 3.86 & 3.09 & \textbf{0.90} & \textbf{0.93} & \textbf{0.80} & 7.50 \\
Cat & 3.12 & 3.48 & 3.89 & 3.09 & \textbf{0.90} & \textbf{0.93} & \textbf{0.80} & \textbf{7.49} \\
CA & 3.13 & 3.50 & 3.87 & 3.08 & 0.89 & \textbf{0.93} & 0.79 & 7.78 \\
FiLM & 3.10 & 3.47 & 3.85 & 3.06 & \textbf{0.90} & \textbf{0.93} & \textbf{0.80} & 7.58 \\ \bottomrule
\end{tabular}
\caption{Evaluation of denoising performance under different acoustic-conditioned schemes. ``Add", ``Cat", and ``CA" refer to addition, concatenation, and cross-attention, respectively.}
\label{tab:acoustic_full}
\end{table*}

The details of the four investigated schemes are as follows:
\begin{itemize}
    \item \textbf{Addition}: The acoustic representation is first linearly projected to align the feature space with the phonetic representation, after which the two are fused via element-wise summation.
    \item \textbf{Concatenation}: The acoustic representation is also projected first by a linear layer, and then the two representations are concatenated along the feature dimension.
    \item \textbf{Cross-attention}: A transformer decoder \cite{Transformer} block is utilized to fuse the two information streams, where the acoustic representation serves as the query and the phonetic one acts as the key and value. The number of attention heads is set to 8.
    \item \textbf{FiLM}: A FiLM \cite{FiLM} module is employed to modulate the phonetic representation by affine transformations, whose scaling and bias parameters are obtained by two separate linear layers.
\end{itemize}
The full metric results, consistent with those reported in the main paper, are presented in Table~\ref{tab:acoustic_full}.

\section{Baselines}
\label{app:baselines}
The details of the baseline systems are as follows:
\begin{itemize}
    \item \textbf{TF-GridNet} \cite{TF-GridNet}: A cutting-edge discriminative model that integrates full- and sub-band modeling in the time-frequency domain. We retrain the model using the official implementation\footnote{\url{https://github.com/espnet/espnet/blob/master/espnet2/enh/separator/tfgridnetv3_separator.py}}. The model consists of 5 blocks with an embedding dimension of 48. Within each TF-GridNet block, both the time and frequency BLSTMs use 100 hidden units per direction. The unfolding operations use a kernel size of 4 and a stride of 1. The self-attention module employs 8 attention heads, with the number of output channels set to 4. 

    \item \textbf{StoRM} \cite{StoRM}: A diffusion-based model that leverages a stochastic regeneration strategy, where a predictive model guides the diffusion process to mitigate artifacts such as vocalizations and breathing noises. We directly utilize the officially released checkpoints\footnote{\url{https://github.com/sp-uhh/storm}} for evaluation on our test sets, with the number of reverse steps set to 30.

    \item \textbf{FlowSE} \cite{FlowSE}: A flow-matching-based model that learns a direct transformation from noisy to clean speech. For a fair comparison, we employ the officially released checkpoint\footnote{\url{https://github.com/Honee-W/FlowSE}}, utilizing the text-free variant with its default 32 sampling steps. It is important to note, as indicated by the authors, that this checkpoint was trained on a smaller dataset than that reported in their paper, which may impact its performance.

    \item \textbf{LLaSE-G1} \cite{LLaSE-G1}: An LM-based model that leverages continuous acoustic representations as inputs to enhance acoustic consistency and generates speech tokens through parallel prediction. We use the official code and checkpoints\footnote{\url{https://github.com/Kevin-naticl/LLaSE-G1}} for evaluation.

    \item \textbf{Adobe Enhance Speech V2}: A cutting-edge commercial model designed to improve the clarity and quality of audio recordings. Inference is performed via its official online API\footnote{\url{https://podcast.adobe.com/en/enhance}} with the default enhancement strength. To facilitate processing, all test utterances are concatenated into several long audio files prior to inference. The enhanced outputs are then segmented back into individual utterances based on their original durations. Since AES-V2 generates 48-kHz outputs, we downsample them to 16 kHz to ensure a fair comparison. 
    
\end{itemize}

\section{Orthogonality Analysis of WavLM Representations}
\label{app:orthogonality_analysis}
\begin{figure}[t]
  \centering
  \centerline{\includegraphics[width=1\columnwidth]{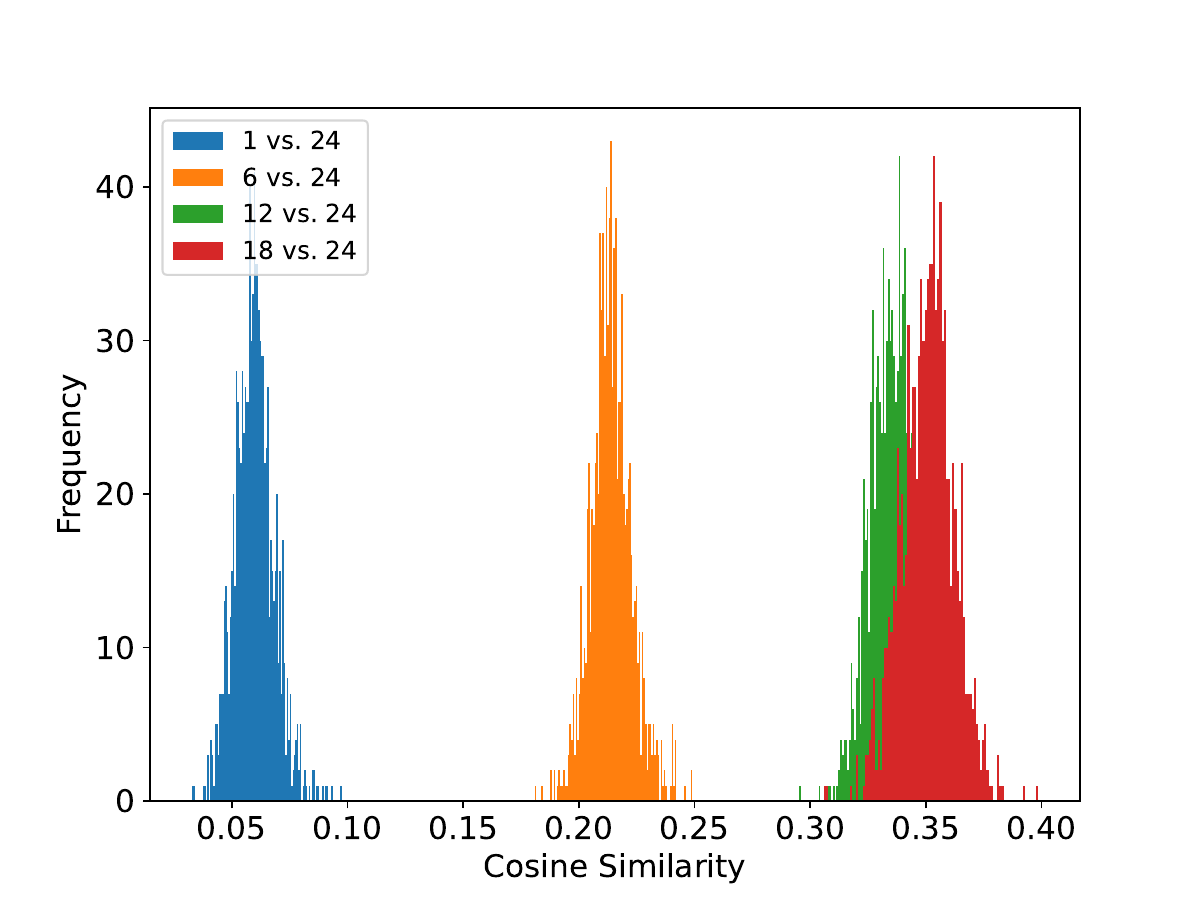}}
  \caption{Orthogonality analysis for different layers of WavLM representations.}
  \label{fig:cossim}
\end{figure}

To investigate the orthogonality of representations across different layers of WavLM, we compute the average cosine similarity between outputs from selected transformer layers on the simulated LibriTTS test set. Specifically, we compare layers 1, 6, 12, and 18 against the final layer (layer 24).
As shown in Fig.~\ref{fig:cossim}, cosine similarity steadily increases with layer depth. Notably, the similarity between layer 1 and layer 24 is nearly zero, with a mean of 0.060 and a standard deviation of 0.0086—corresponding to an angle of approximately $\textrm{86}^\circ
$. This suggests that the representations from layer 1 and layer 24 are almost orthogonal in the embedding space, underscoring the effectiveness of our additive acoustic-conditioned scheme.

\begin{figure*}[!htbp]
  \centering
  \centerline{\includegraphics[width=0.75\linewidth]{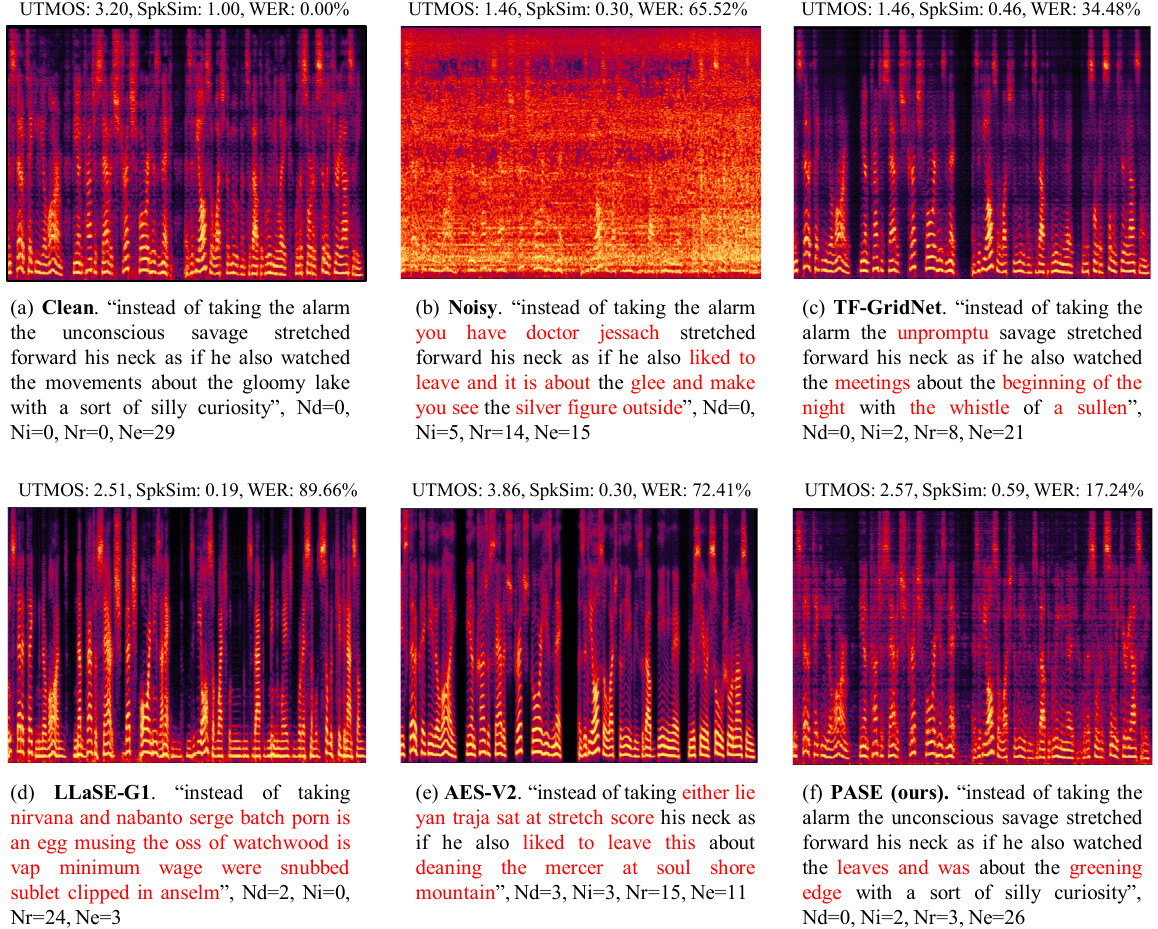}}
  \caption{An audio example demonstrating PASE’s superiority in maintaining linguistic accuracy.}
  \label{fig:example1}
\end{figure*}

\begin{figure*}[!htbp]
  \centering
  \centerline{\includegraphics[width=0.75\linewidth]{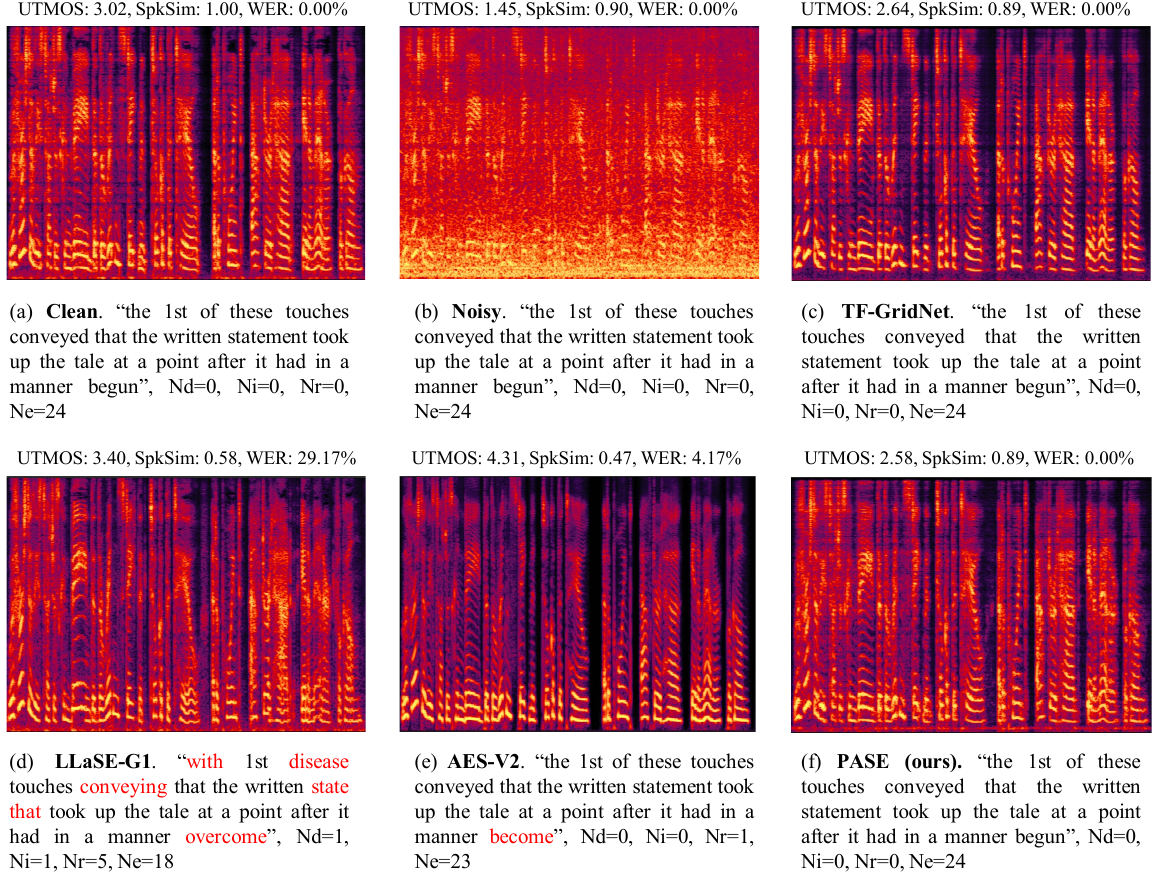}}
  \caption{An audio example demonstrating PASE’s superiority in preserving speaker characteristics.}
  \label{fig:example2}
\end{figure*}

\begin{figure*}[!htbp]
  \centering
  \centerline{\includegraphics[width=0.75\linewidth]{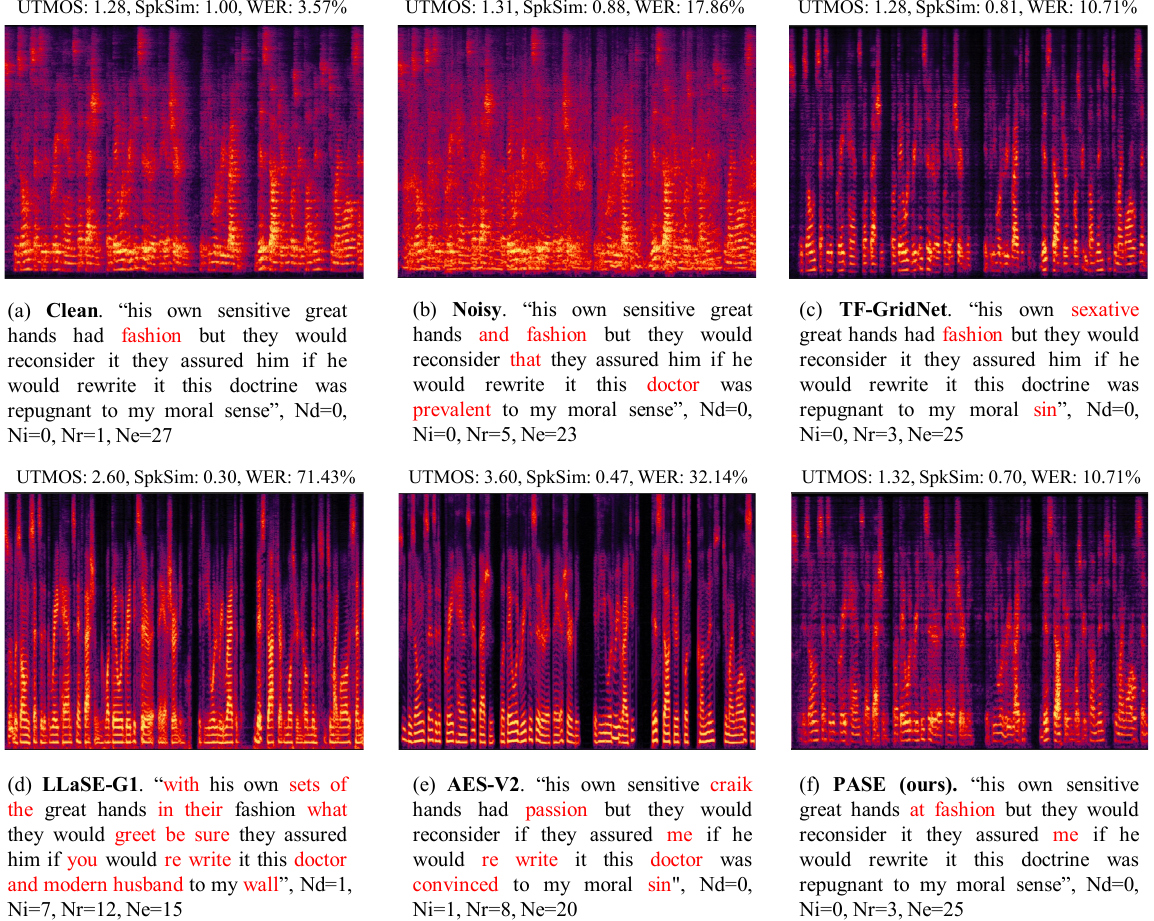}}
  \caption{An audio example demonstrating how severe reverberation can lead to anomalously low UTMOS scores, even when the perceived audio quality remains high.}
  \label{fig:example3}
\end{figure*}

\section{Audio Examples Visualization}
\label{app:audio_examples}
We present three groups of audio examples, each comprising six utterances: clean speech, noisy speech, and enhanced outputs from TF-GridNet, LLaSE-G1, AES-V2, and our proposed PASE. 
We omit comparisons with FlowSE and StoRM here, as their performance is relatively poor.
The first two groups, drawn from the LibriTTS simulated test set, highlight PASE’s superiority in preserving linguistic integrity and speaker characteristics, respectively. The third group, from the \textit{with-reverb} DNS1 test set, illustrates how dereverberation can lead to abnormally low UTMOS scores, despite high perceptual quality.
For each utterance, we display its UTMOS, SpkSim, and WER scores above the spectrogram, along with the ASR transcript below. The ASR output is accompanied by word-level error statistics: deletions (Nd), insertions (Ni), replacements (Nr), and equal words (Ne). Notably, as DNS1 does not provide reference transcripts, we manually annotated this sample.

As shown in Fig.~\ref{fig:example1}, when the input is heavily corrupted by noise, the discriminative TF-GridNet yields perceptually degraded outputs with only moderate linguistic accuracy. Generative models like LLaSE-G1 and AES-V2 suffer from even more severe hallucinations, reflected in low SpkSim and high WER. In contrast, PASE achieves the best WER and SpkSim, while also maintaining a relatively high UTMOS—highlighting its effectiveness in suppressing hallucinations and preserving overall quality.

Fig.~\ref{fig:example2} illustrates that when the noisy input exhibits a moderately high SNR, all models—except LLaSE-G1—are capable of producing semantically accurate enhanced outputs. Notably, while the generative model AES-V2 achieves a high UTMOS, its comparatively low SpkSim suggests a limited ability to preserve speaker characteristics. In contrast, PASE attains a more moderate UTMOS—still indicative of satisfactory perceptual quality—while maintaining a high SpkSim of 0.89, demonstrating a well-balanced trade-off between speech quality and speaker fidelity.

In Fig.~\ref{fig:example3}, we present a noisy speech sample with strong reverberation. While LLaSE-G1 and AES-V2 generate “dry” and perceptually clean outputs, they still exhibit both acoustic and linguistic hallucinations, as evidenced by their low SpkSim and poor WER scores. In contrast, our PASE effectively removes the reverberation; however, the dereverberation process appears to induce an abnormally low UTMOS score, despite the high perceptual quality. Nevertheless, PASE consistently maintains high SpkSim and low WER, demonstrating its robustness under severe reverberant conditions.

\end{document}